%% file: Paper.tex
\documentclass[amsmath,twocolumn,superscriptaddress]{revtex4-1}
\usepackage{epsfig}
\usepackage{amssymb}
\usepackage{amsmath}
\usepackage{amsfonts}
\usepackage{braket}
\usepackage[T1]{fontenc}
\usepackage{bm}
\usepackage{graphicx}
\usepackage{xcolor}
\usepackage{verbatim}
\usepackage{soul}
\usepackage{tikz}
\usetikzlibrary{matrix,shapes,arrows,positioning,chains}
\usetikzlibrary{shapes.geometric}
\usetikzlibrary{shapes.arrows}
\usepackage{array}
\usepackage{enumitem}
\usepackage{color}
\usepackage{physics}
\usepackage[export]{adjustbox}
\usepackage[percent]{overpic}
\DeclareMathOperator*{\argmax}{argmax}

\begin{document}

\title{Simulating groundstate and dynamical quantum phase transitions on a superconducting quantum computer}

\author{James Dborin}
\affiliation{London Centre for Nanotechnology, University College London, Gordon St., London, WC1H 0AH, United Kingdom}

\author{Vinul Wimalaweera}
\affiliation{London Centre for Nanotechnology, University College London, Gordon St., London, WC1H 0AH, United Kingdom}

\author{F. Barratt}
\affiliation{Department of Physics, University of Massachusetts, Amherst, MA  01003, USA }

\author{Eric Ostby}
\affiliation{Google Quantum AI, 80636 Munich, Germany}

\author{Thomas E. O'Brien}
\affiliation{Google Quantum AI, 80636 Munich, Germany}

\author{A.~G. Green}
\affiliation{London Centre for Nanotechnology, University College London, Gordon St., London, WC1H 0AH, United Kingdom}
\affiliation{email: andrew.green@ucl.ac.uk}

\date{\today}
\begin{abstract}
    We optimise a translationally invariant, sequential quantum circuit on a superconducting quantum device to simulate the groundstate of the quantum Ising model through its quantum critical point. We further demonstrate how the dynamical quantum critical point found in quenches of this model across its quantum critical point can be simulated. Our approach avoids finite-size scaling effects by using sequential quantum circuits inspired by infinite matrix product states. We provide efficient circuits and a variety of error mitigation strategies to implement, optimise and time-evolve these states. 
\end{abstract}
\maketitle
\tableofcontents

\vspace{-0.2in}

\section{Introduction} 
\label{sec:introduction}
\input{Introduction.tex}

\section{Results}
\input{Results.tex}

\section{Discussion}
 \input{Discussion.tex}

\section{Methods}
\label{app:Methods}
\input{MethodsA.tex}
\input{MethodsB.tex}


\bibliographystyle{naturemag}
\bibliography{Bibliography.bib}

\end{document}



\title{Supplementary Materials: Simulating groundstate and dynamical quantum phase transitions on a superconducting quantum computer}

\author{James Dborin}
\affiliation{London Centre for Nanotechnology, University College London, Gordon St., London, WC1H 0AH, United Kingdom}

\author{Vinul Wimalaweera}
\affiliation{London Centre for Nanotechnology, University College London, Gordon St., London, WC1H 0AH, United Kingdom}

\author{F. Barratt}
\affiliation{Department of Physics, University of Massachusetts, Amherst, MA  01003, USA }

\author{Eric Ostby}
\affiliation{Google Quantum AI, 80636 Munich, Germany}

\author{Thomas E. O'Brien}
\affiliation{Google Quantum AI, 80636 Munich, Germany}

\author{A.~G. Green}
\affiliation{London Centre for Nanotechnology, University College London, Gordon St., London, WC1H 0AH, United Kingdom}
\affiliation{email: andrew.green@ucl.ac.uk}

\date{\today}
\begin{abstract}
In this supplementary material we further explain the methods used to construct circuits for time-evolution. We pay particular attention to the tradeoffs between circuit approximations and fidelity in constructing cost-functions.
\end{abstract}
\maketitle

\section{Order of Trotterisation}
\begin{figure*}[t]
\includegraphics[width=0.75\textwidth]{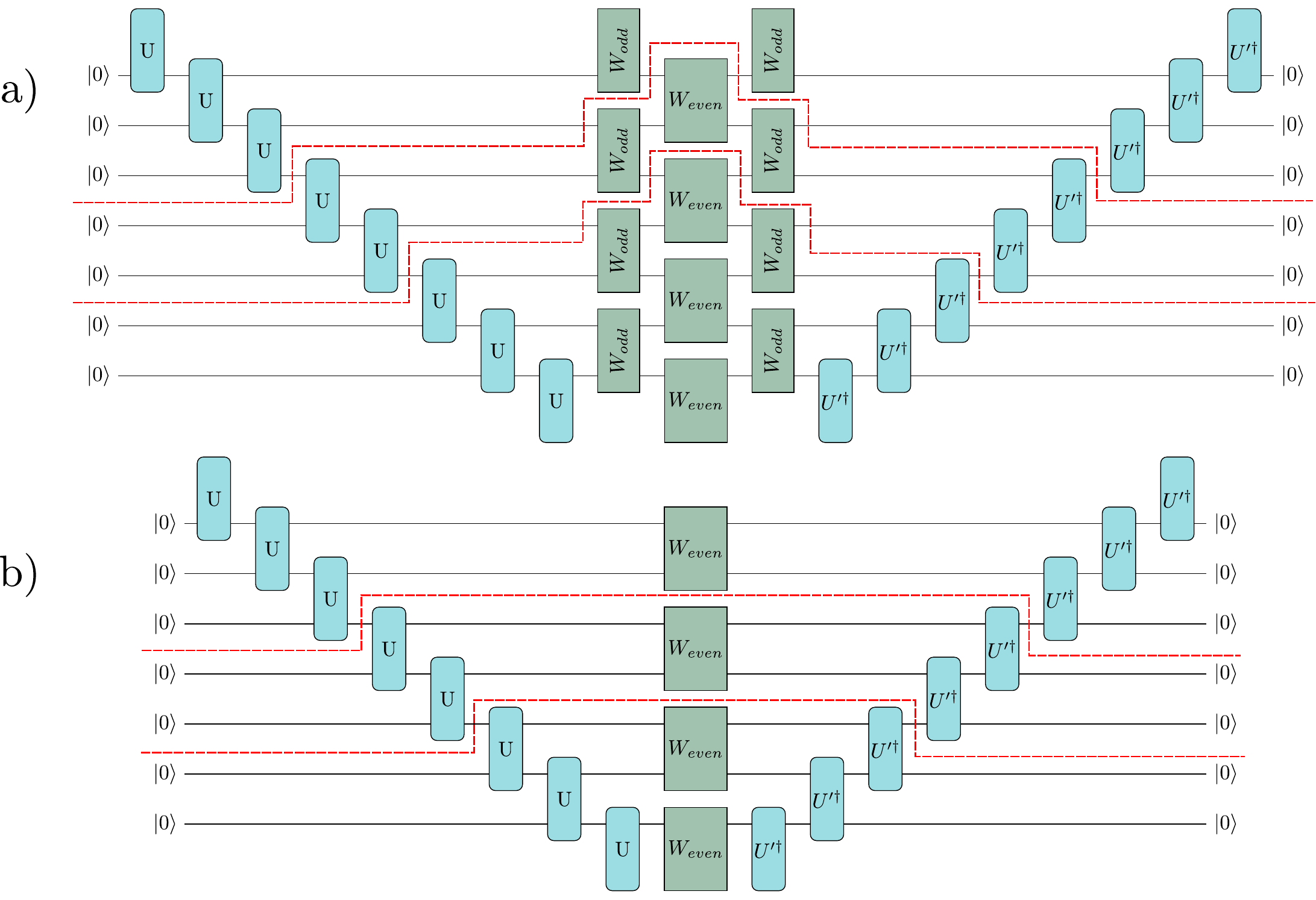}
\caption{{\bf Time-evolution Transfer Matrix} a) The overlap $|\langle \psi(U') | e^{{\cal H} dt} | \psi(U) \rangle |^2$, calculated with a second order Trotterisation. Here $W_{even}=e^{i{\cal H}_e dt}$ and $W_{odd}=e^{i{\cal H}_o dt/2}$ . This overlap is formally an infinite width and depth circuit. b) An expression for the overlap with one half timestep as used here. Mapping back to a translationally-invarient variational manifold increase sthe effective order of the time-integration.}
\label{fig:Trotterisation}
\end{figure*}

One of the key refinement parameters in time-evolving circuit states is the order of Trotterisation. Supplementary Fig.\ref{fig:Trotterisation} illustrates the different transfer matrices that result from a second order Trotterisation and from the half-order Trotterisation.  Because of the way in which the states are projected back to translationally invariant states, this update is in fact found to be effective to higher order in $dt$. The results presented below and in the main paper are exclusively for the half-order Trotterisation in order to avoid the deeper circuit required for the second order Trotterisation.

\section{Cost Functions for Time-evolution}
As discussed in Section II C of the main text, finding an appropriate cost-function for time-evolution quantum circuit iMPS involves a subtle trade-off between the analytical approximation to the eigenvalue of the transfer matrix - determined by order of Trotterisation, order of power method used, and accuracy of approximations to the transfer matrix fixed-points - and the infidelities of representing this circuit on a real device. Here we give results for a number of different cost functions, the dynamics that they predict in the absence of circuit noise and their realisation on the Rainbow device.

\subsection{Approximating the right fixed point with $|0 \rangle \langle 0|$}
\begin{figure*}[t]
\includegraphics[width=0.8\textwidth]{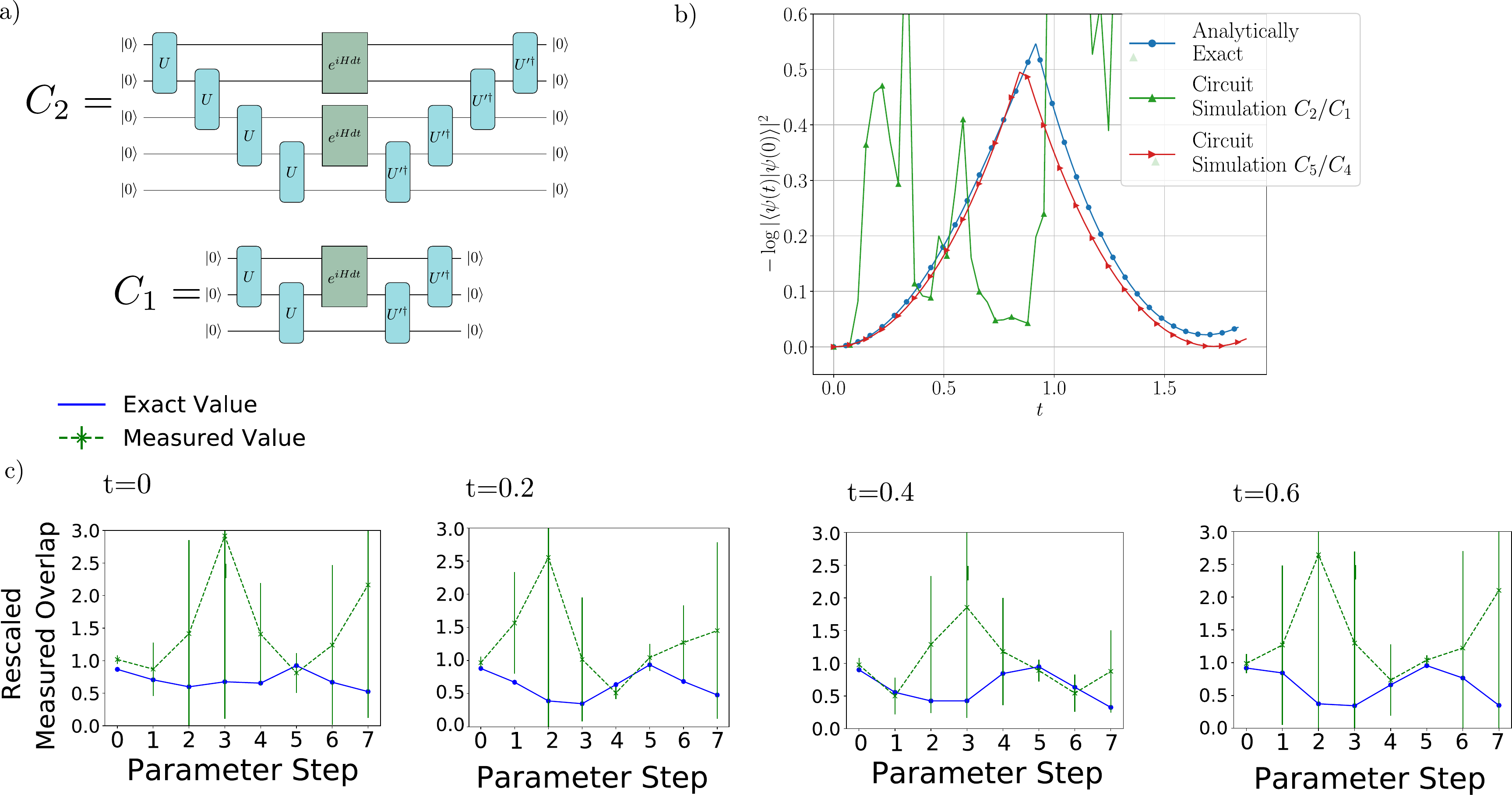}
\caption{{\bf Time-Evolution Circuits and Results: approximating the right fixed point  with} $|0\rangle  \langle 0|$ {\bf Ratio of} $C_2/C_1 \approx \lambda$.
a) {\it Time-evolution circuit:} This circuit approximates the left fixed point as ${\bm I}$ and the right fixed point as   $|0\rangle  \langle 0|$. The probability of measuring $|0 \rangle^{\otimes N}$ at the output is divided by the appropriate Loschmidt echo and the ratio of the output with two copies of the transfer matrix divided by the output for a circuit with one copy of the transfer matrix. This gives an approximation to $\lambda$ --- the principal eigenvalue of the transfer matrix.
b) {\it Dynamical Quantum Phase Transition in the Quantum Ising Model:}   Exact analytical results of the time evolution and a simulation using direct stochastic optimisation of the circuit in a) in the absence of noise. Taking the ratio of $C_2/C_1$ approximated with the circuit shown in a) does not reproduce the correct time evolution. Calculating $C_5/C_4$ using the same approximation for the right fixed point does a much better job at capturing the time-evolution suggesting that it is the low order of the power method that leads to the failure.
c){\it Cost-function evaluated on Google's Rainbow Device:} The cost function evaluated along a linear interpolation from $U'=U$ through the optimal value of $U'$ as in Fig.3 of the main text. Though this displays a peak at the optimum value it is not the global optimum of the circuit.}
\label{fig:Sup1}
\end{figure*}

\begin{figure*}[b]
\includegraphics[width=0.8\textwidth]{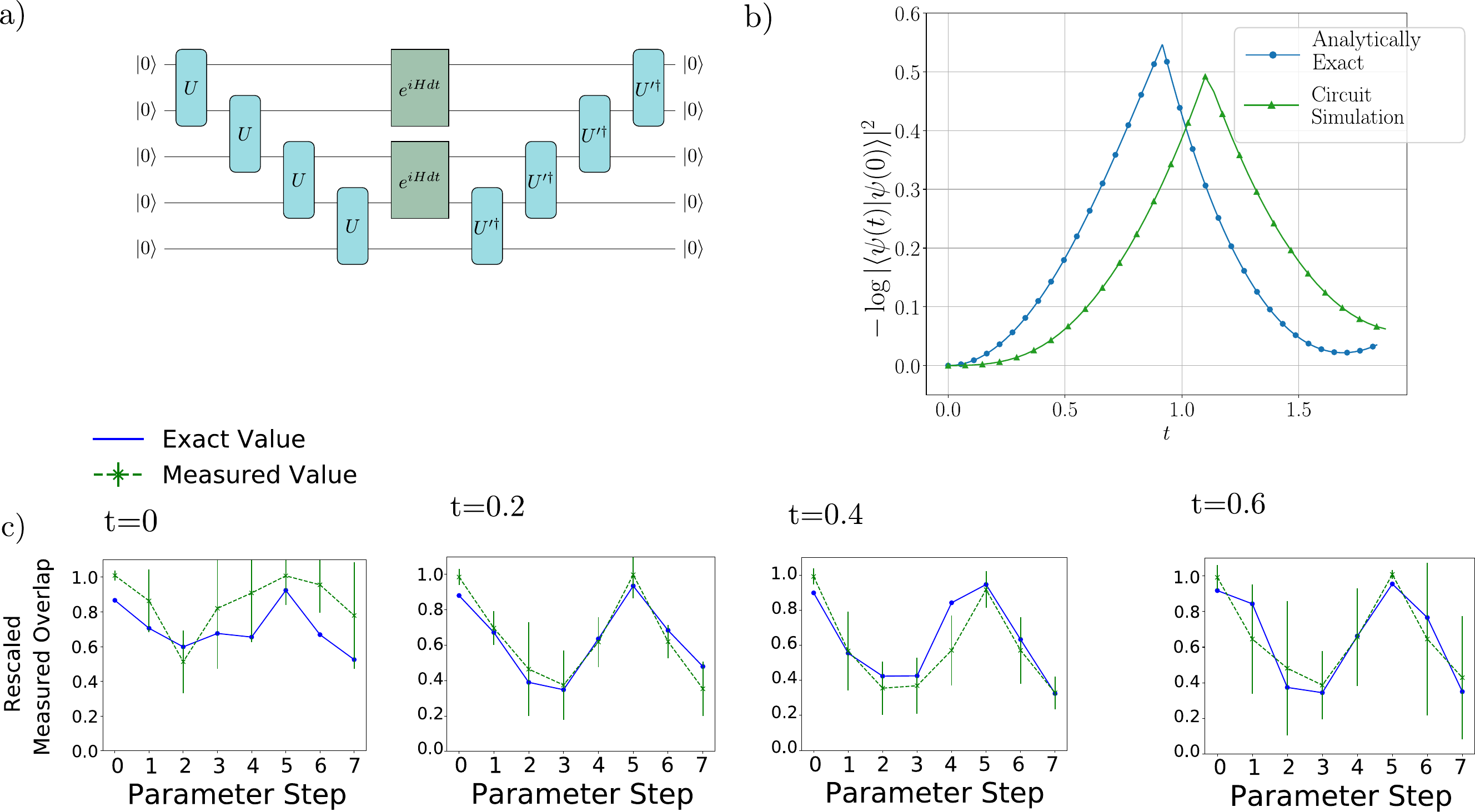}
\caption{{\bf Time-Evolution Circuits and Results: approximating the right fixed point  with} $|0\rangle  \langle 0|$  $C_2 \approx \lambda^2$.
a) {\it Time-evolution circuit:} Here we use the first of the two circuits used in Supplementary Fig. \ref{fig:Sup1}. The probability of measuring $|0 \rangle^{\otimes N}$ at the output is divided by the appropriate Loschmidt echo. This gives an approximation to $\lambda^2$ --- the square of the principal eigenvalue of the transfer matrix.
b) {\it Dynamical Quantum Phase Transition in the Quantum Ising Model:} Exact analytical results of the time evolution and from a simulation using direct stochastic optimisation of the circuit in a) in the absence of noise. Surprisingly, although the same circuit is used as in Supplementary Fig. 1, in this case, the circuit captures key features of the evolution when calculated in the absence of noise.
c){\it Cost-function evaluated on Google's Rainbow Device:} The cost function evaluated along a linear interpolation between the initial and optimal parameters as in Fig.3 of the main text. Although the cost function measured on the Rainbow deice does track its calculation in the absence of noise, we find that the peak does not always occur at the optimal value.  }
\label{fig:Sup2}
\end{figure*}

First we show in Figs. \ref{fig:Sup1} and \ref{fig:Sup2} the result of using an identity approximation to the left eigenvalue and a simple $| 0 \rangle \langle 0|$ approximation to the right eigenvalue. As shown in \ref{fig:Sup1}, this performs very poorly when using the ratio $C_2/C_1$ to estimate the principle eigenvalue of the transfer matrix. This is true even when the circuits are calculated in simulations in the absence of noise. This is apparently due to the low order of the power method, despite the good approximation to the left fixed point. When the same method is used to calculate $C_5/C_4$ the approximation performs well in simulations without noise.

Curiously, simply using $C_2$ as an approximation to $\lambda^2$ performs much better. Supplementary Fig.\ref{fig:Sup2} shows that, in the absence of circuit errors, this circuit does a passable job of capturing the dynamical quantum phase transition in the quench dynamics of the quantum Ising model. However, when implemented on the Rainbow device, the optimum measured cost function is not at the correct updated values - even after we carry out our rescaling to account for depolarisation error.

%

\subsection{Improved approximations to the right fixed-point. }
While in principle, a good approximation to the principal eigenvalue of the transfer matrix can be found using a good initial approximation to only either the left or right fixed-points of the transfer matrix. We find in practice that we get much better results if our approximations to both left and right fixed points are good. Supplementary Figs.\ref{fig:Sup3} and \ref{fig:Sup4} show the results of progressive improvements to the approximation.

In Supplementary Fig. \ref{fig:Sup3} we show the results of using the same circuit as in Fig. 3 in the main paper, but measuring the probability of $|0 \rangle^{\otimes 6}$ at the output rather than post-selecting the two top right qubits on $|0 \rangle$. This amounts to using $U$ and $U'$ to construct an approximation to the right fixed-point. The resulting circuit does a passable job of capturing the dynamical quantum phase transition in the absence of circuit errors, and the cost function is apparently tracked rather well when implemented on the Rainbow device. However, the measured optimum value of this cost function is not correct. Post-selecting on the top right qubits corrects these deficiencies. This procedure factors out the $U'$-dependent norm of the approximation to the right fixed-point, giving better performance both under simulation in the absence of error and when implemented on the Rainbow device.

Finally, we report on two further approximations to the right fixed point that do not require post-selection, and may perform even better than the circuit reported in the main paper. Unfortunately, it was not possible to test these circuits on the Rainbow device as they were discovered after the machine was taken offline. These correspond to approximating the right fixed-point by that obtained when overlapping the state described by $U$ with itself. This is a good approximation since evolving the state described by $U'$ backwards in time ought to recover the state described by $U$. The two approximation shown in Supplementary Fig.\ref{fig:Sup4} correspond to using the fixed point determined by solving the fixed-point equations of Fig.4c in the main paper for $V(U)$ and an approximation to this obtained by simply overlapping $U$ with itself to construct an approximation to the right fixed point. Both approximations do an excellent job of capturing the evolution in the absence of noise and we believe will provide a good starting point for future investigation.

\begin{figure*}[ht]
\includegraphics[width=0.8\textwidth]{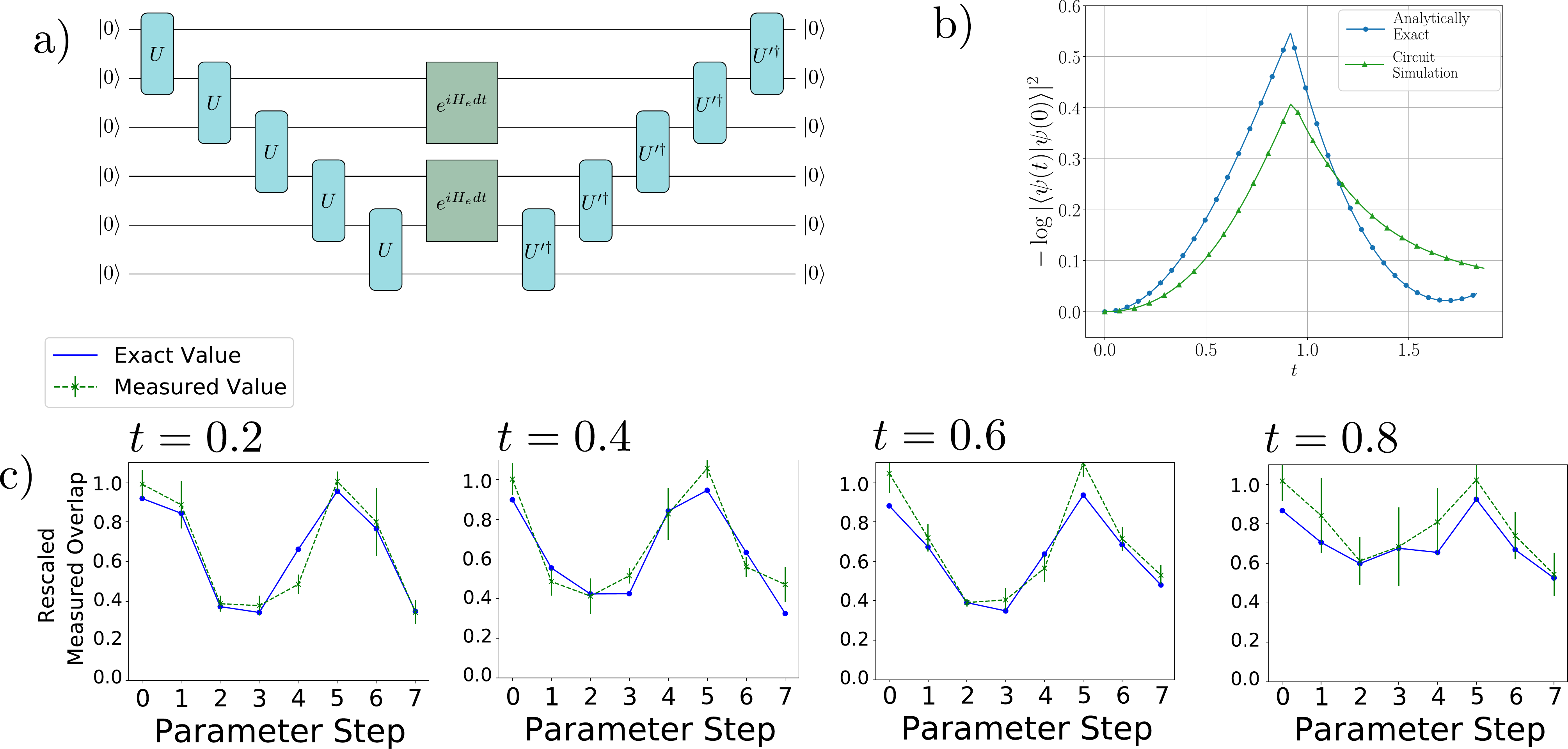}
\caption{{\bf Time-Evolution Circuits and Results: Improved Approximation to right fixed-point using} $U$ {\bf and} $U'$ {\bf without post selection} :
a) {\it Time-evolution circuit:} Here we use the same circuit as shown in Fig 3. in the main paper, except we evaluate the full overlap through the probability of measuring $|0 \rangle^N$ on all outputs. Fig. 3 of the main paper post-selected the final two qubits on $|0 \rangle$. This latter procedure allows for the non-unit norm of the resulting approximation to the right fixed point and gives better results.
b) {\it Dynamical Quantum Phase Transition in the Quantum Ising Model:}  Exact analytical results of the time evolution and from a simulation using direct stochastic optimisation of the circuit in a) in the absence of noise.
While some features of the dynamical phase transition are captured, the result is not as good as obtained using post-selection in the main text.
c){\it Cost-function evaluated on Google's Rainbow Device:} The cost function evaluated along a linear interpolation between the initial parameters and the optimal update as in Fig.3 of the main text. Again, although the measured values track the calculated values in the absence of noise, the optimal value is not in the correct place. }
\label{fig:Sup3}
\end{figure*}
%
\begin{figure*}[htb]
\includegraphics[width=0.8\textwidth]{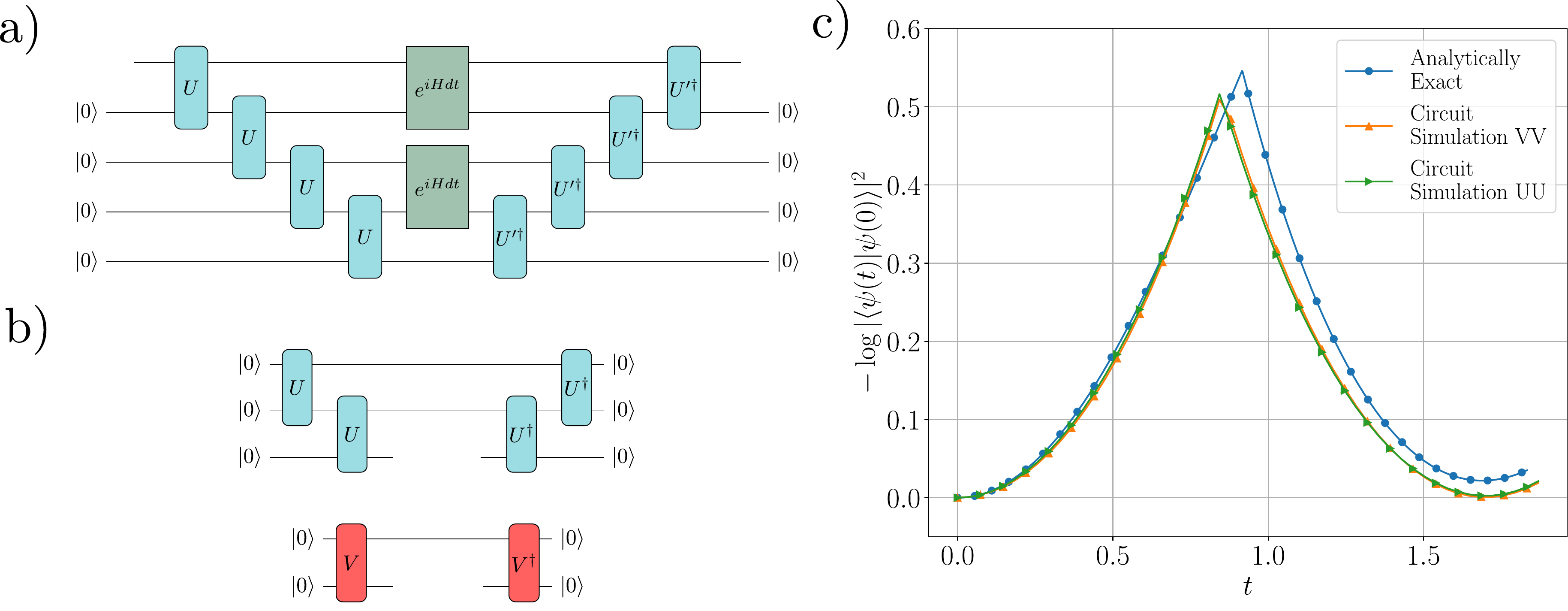}
\caption{{\bf Time-Evolution Circuits and Results: Improved approximation to right fixed-point using } $U$ {\bf and} $U$, {\bf and} $V$:
a) {\it Time-evolution circuit:} Here we use two different cost functions corresponding to the two approximations to the right fixed point shown in b). The probability of measuring $|0 \rangle^{\otimes N}$ at the output gives an approximation to $\lambda^2$.
b) {\it Approximation to right fixed point:} The right fixed point is approximated in two ways. First using  $U$ alone. In this case, the norm of this approximation to the right fixed point is automatically $1$ so post-selection is not required. Secondly we use $V$ as evaluated in the energy optimisation. This in principle gives a better approximation to the right fixed point, but involves solving the fixed point equations.
c) {\it Dynamical Quantum Phase Transition in the Quantum Ising Model:} Exact analytical results of the time evolution and from a simulation using direct stochastic optimisation of the circuits in a) in the absence of noise. The results are very good using either $U$ alone or using $V$. Since the latter requires stochastic optmisation of the fixed-point equations, we propose the approximation using $U$ alone as a starting point for future investigation. }
\label{fig:Sup4}
\end{figure*}

\bibliographystyle{naturemag}
\bibliography{Bibliography.bib}

%% file: Introduction.tex
The simulation of chemical reactions and materials properties, and the discovery of new pharmaceutical compounds are anticipated to be major applications of quantum computers. The rapid development of various approaches to constructing NISQ devices brings simple models of these use cases within the realm of possibility. Finding challenging but feasible problems that can be implemented on current devices is crucial. As well as demonstrating the progress that has been made, these serve to highlight required improvements and, ideally, have the possibility of quantum advantage when suitably powerful quantum computers are developed. Many problems that fit this bill are to be found in condensed matter. They can be scaled to fit current machines while retaining scientific and technological relevance.

Strongly correlated condensed matter systems are amongst those most likely to yield a quantum advantage\cite{arute2020observation,babbush2019quantum,stanisic2021observing,tazhigulov2022simulating,satzinger2021realizing, chiaro2022direct}. These are systems in which the underlying spins or electrons are strongly renormalised and whose quantum properties are beyond the reach of standard perturbative or density functional type approaches. Quantum criticality\cite{sachdev2000quantum,coleman2005quantum,sachdev2011quantum} is one of the few collective organising principles that has been able to make sense of a large class of strongly correlated phenomena. The term --- first coined by Hertz\cite{hertz1976quantum} --- refers to systems in the vicinity of a zero-temperature phase transition driven by quantum fluctuations. Such systems display universal spatial and temporal correlations that are not seen in purely classical problems and can underpin the formation of entirely new quantum phases. Diverging correlation lengths near the quantum critical point make these states challenging for numerics, driving the development of state of the art tools such as dynamical mean field theory\cite{parcollet2004cluster} (to deal with the Mott transition), singlet Monte Carlo\cite{sandvik2007evidence} (to deal with deconfined quantum criticality) and tensor networks amongst many others.

\begin{figure*}[t]
\includegraphics[width=0.78\textwidth]{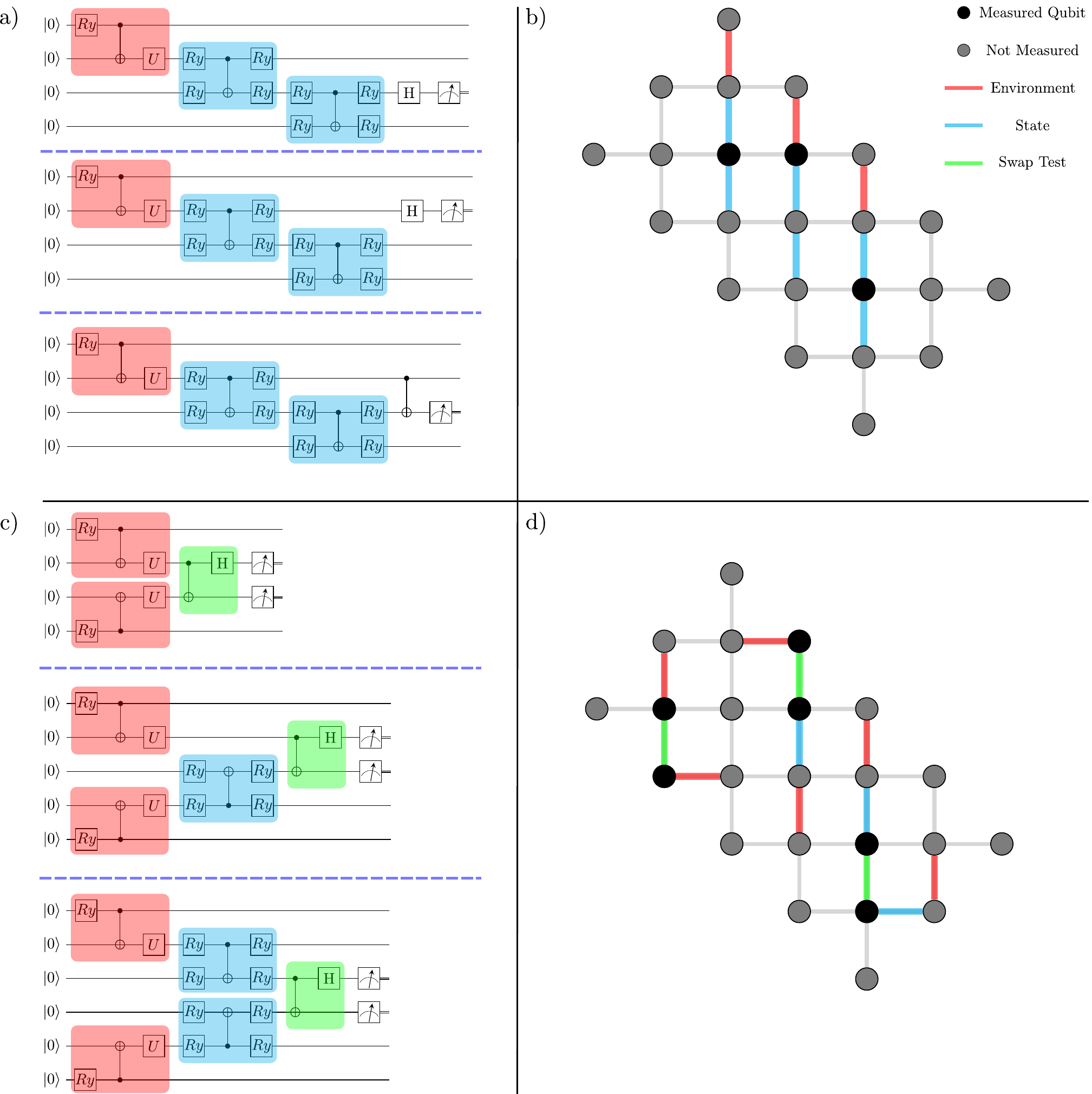}
\caption{{\bf Translationally Invariant Circuit Matrix Product States for groundstate optimisation}. a) iMPS circuits to calculate the three terms in the Ising Hamiltonian. The blue boxes indicate realisations of the state unitary $U$ with its factorisation to the Rainbow gate set. The red boxes highlight the environment tensor $V$. b) Example of the Ising circuits laid out on the Rainbow device. The best performing qubits were chosen on each day of experiment. c) iMPS Circuits to to determine $V\equiv V(U)$. The three elements shown comprise the the three elements required to compute the trace distance between the left- and right-hand sides of the fixed-point equations shown in Fig.\ref{fig:FixedPointCircuits}. The gates needed to perform the required swap test are highlighted in green. d) Example of the trace distance circuits laid out on the rainbow device. 
}
\label{fig:GroundstateCircuits}
\end{figure*}

As well as providing the best classical numerical approach for 
many
spin systems, tensor network methods can be directly translated to quantum circuits\cite{schon2005sequential,circuits_are_mps,Smith2022crossing,Lin2021real,barratt2021parallel}. Doing so has potential quantum advantage over the classical implementations\cite{Lin2021real}, and tensor networks allow initialisation and improvement of quantum circuits in a way that can circumvent the barren plateaux that potentially plague quantum circuits. 
There are particular advantages in applying these methods to describe quantum critical systems. Diverging correlation lengths lead to strong finite-size effects which can be avoided with tensor networks by working directly in the thermodynamic limit. There may be advantages in combining tensor network methods with machine learning tools\cite{carrasquilla2017machine,stoudenmire2016supervised,huggins2019towards} to extract simulation results as used recently in classical numerics. Moreover, there is potentially excellent fit between tensor network simulation methods and matrix product operator-based error mitigation\cite{guo2022quantum,herrmann2021realizing}.
While we focus on  one-dimensional matrix product states (MPS), reflecting the limitations of current devices, translations of these methods to higher dimensions have been proposed\cite{banuls2008sequentially,zaletel2020isometric}.

\begin{figure*}[t]
\includegraphics[width=0.9\textwidth]{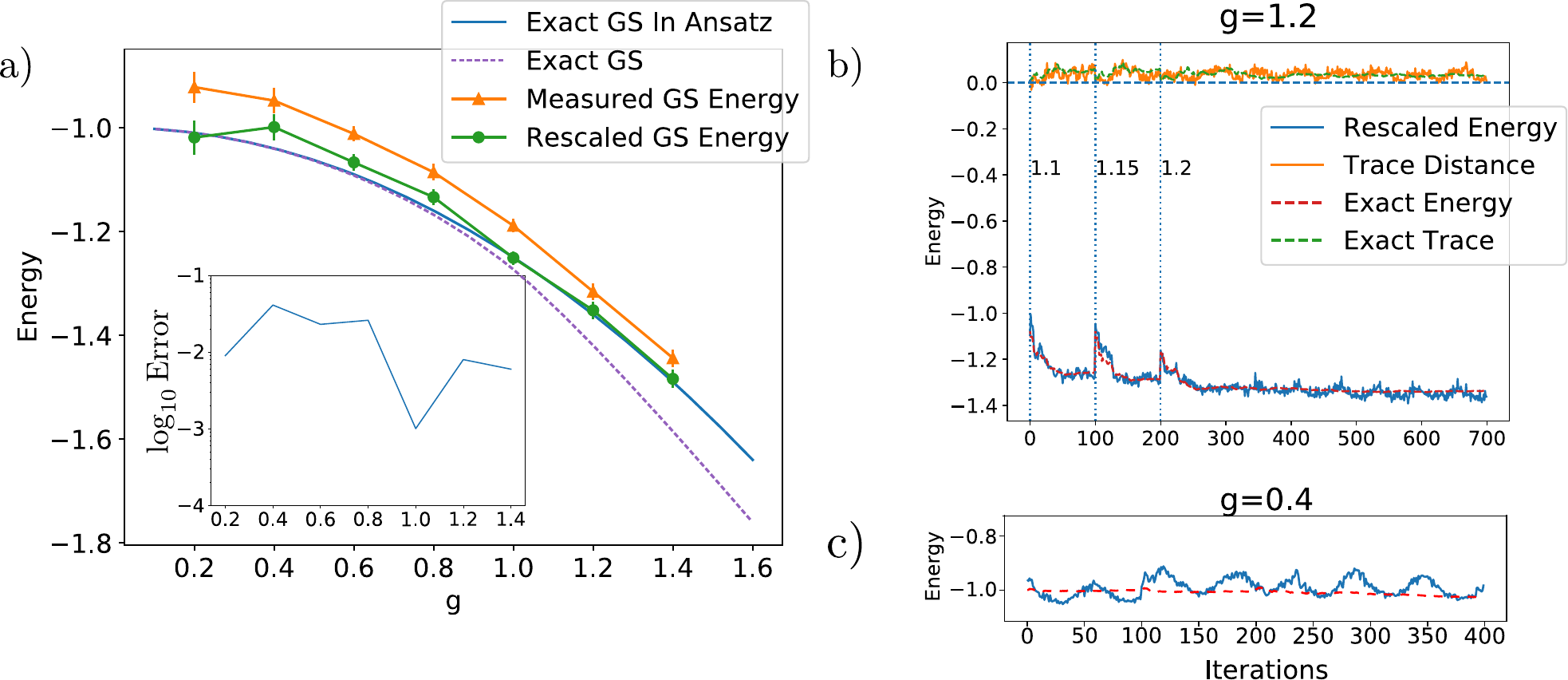}
\caption{{\bf Optimisation of Groundsates of the Quantum Ising Model}. a) The blue curve gives the exact energy of the quantum Ising model optimised over bond order $D=2$ MPS within the ansatz class used on the device. The dashed purple curve gives the analytically exact groundstate energy. The green and orange curves show energies of the optimised on the Rainbow device with and without Loschmidt rescaling to allow for depolarisation, respectively. Measurement errors are corrected using a confusion matrix in both cases. Except for the anomalous point at $g=0.4$ affected by an uncontrolled oscillation on the device (see c), the rescaled results are within $2.2 \%$ of the exact values. This is true even at the quantum critical point, $g=1$. The inset shows the log-deviation in measured energy compared to the exact value for the circuit ansatz.  $J$ has been rescaled to $1$ in these plots. 
b) A typical optimisation curve showing the measured Hamiltonian and the trace-distance. 
The latter is defined in the main text. A value of zero value for it indicates that $U$ and $V$ consistently describe the local properties of translationally invariant state. 
Optimisation is carried out using the simultaneous perturbation stochastic approximation (SPSA) and a quasi-adiabatic change in the value of the transverse field $g$. The optimisation is made in stages changing $g$ by discrete amounts and allowing the optimisation to stabilise before incrementing further. The vertical dashed lines indicate junctures when the transverse field $g$ is altered.   c) Oscillations in measured energy found around $g=0.4$. }
\label{fig:GroundstateResults}
\end{figure*}
%

Here we demonstrate that translationally invariant MPS (iMPS) can be used to simulate quantum critical systems, in the thermodynamic limit, on Google's Rainbow device  --- a quantum device that shares the Sycamore architecture\cite{SupremacyPaper} . We focus upon the quantum Ising model which has a quantum phase transition in its groundstate properties\cite{sachdev2000quantum,sachdev2011quantum} and in its dynamics\cite{pollmann2010dynamics,dynamical_phase_transitions}. 
We show that --- with appropriate error mitigation --- the groundstate of the quantum Ising model can be found with high accuracy even at the quantum critical point. We also demonstrate a cost-function that, by balancing analytical approximations and error mitigation strategies, 
faithfully tracks time-evolution through the dynamical quantum phase transition. The resulting circuits are considerably simpler than previous proposals\cite{barratt2021parallel}

%% file: Results.tex
We begin discussion of our results by introducing the quantum Ising model and its key features. We then show how translationally invariant, quantum circuit iMPS, together with suitable error-mitigation strategies, can be used to determine its groundstate properties. We report the result of applying these methods on Rainbow. Next, we introduce circuits that can be used to simulate the quantum dynamics, and the balance of analytical approximations and error mitigation that permit them to be implemented on the Rainbow device.

\begin{figure*}[t]
\includegraphics[width=0.9\textwidth]{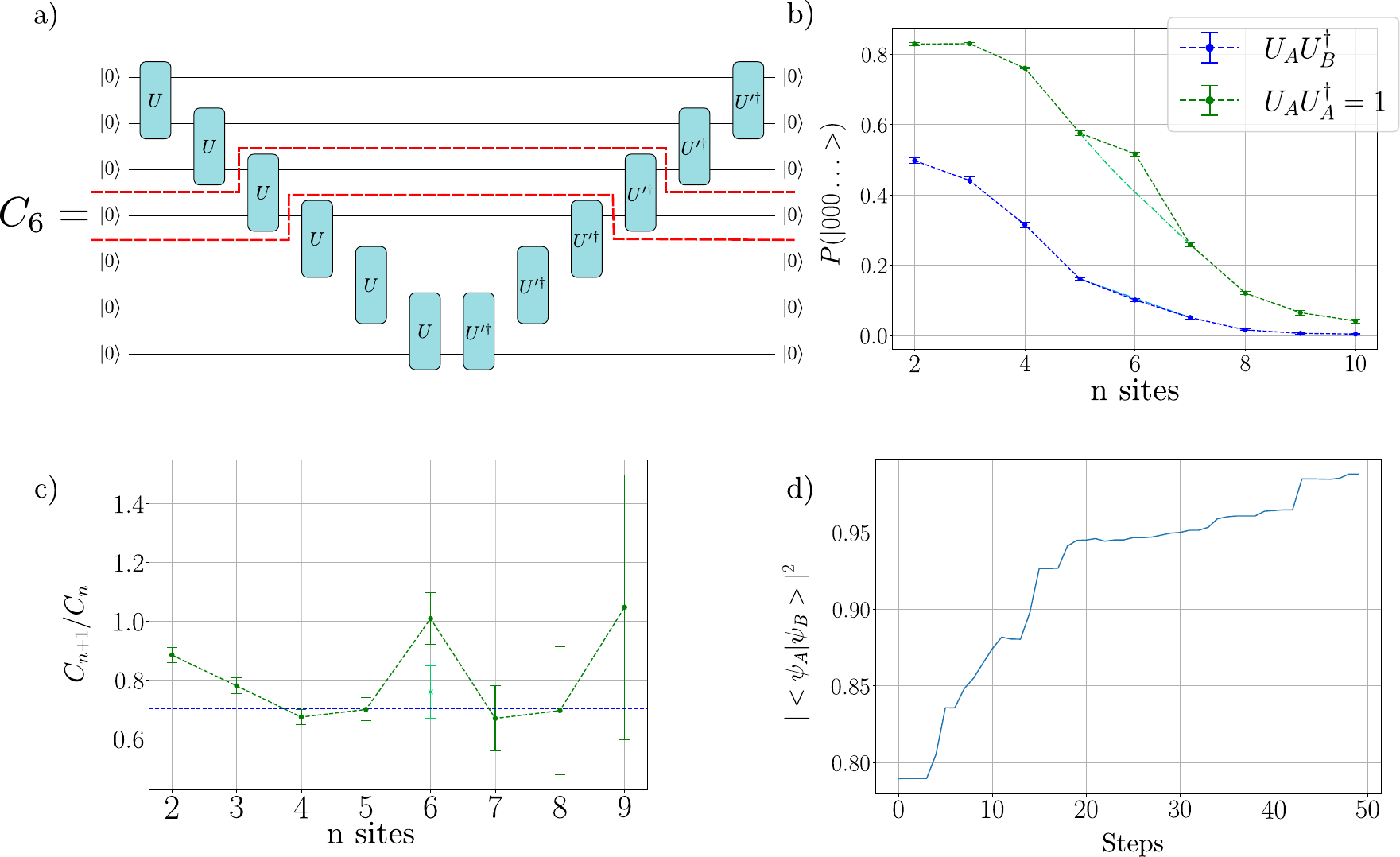}
\caption{{\bf Calculating State Overlaps}.
 a) The overlap of two translationally invariant states parametrised by $U$ and $U'$ is given by $\lim_{n \rightarrow \infty } C_n \rightarrow \lambda^n$, where $\lambda$ is the principle eigenvalue of the transfer matrix delimited by the red dotted line. $C_n$ is evaluated on circuit by measuring the probability of $ | 0 \rangle^{\otimes (n+1)}$ at the output. In order to correct for depolarisation errors, we divide by the Loschmidt echo obtained by evaluating the circuit at $U'=U$.
b) Overlaps $C_n(U_A,U_B)$ and Loschmidt echo $C_n(U_A,U_A)$ evaluated on Rainbow as a function of the order of power method $n$. 
c) The ratio $C_{n+1}/C_n$ obtained from the data in b). The overlap begins to converge as the circuit depth --- measured by $n$ --- rises. By $n=4$ the measured value overlaps within error bars with the exact value of the principal eigenvalue of the transfer matrix (aside from the outlier at $n=6$ which occurs due to an error in the estimate of the Loschmidt echo. This is corrected by the interpolation shown in b) ). The circuit depth increases with $n$ leading to increased error bars. However, the result is still within errors suggesting that useful information can be extracted even from these deeper circuits.
d) A demonstration that stochastic optimisation of $U_B$ using SPSA converges to $U_B=U_A$. }
\label{fig:StateOverlap}
\end{figure*}

 \subsection{Quantum Phase Transitions}
 \label{sec:Problems}
The quantum or transverse field Ising model is one of the best understood models that exhibits quantum critical phenomena.
Its Hamiltonian is given by
\begin{equation}
{\cal H} = \sum_i \left[ J \hat Z_i \hat Z_{i+1} + g \hat X_i \right],
\label{eq:IsingModel}
\end{equation}
where $\hat Z$ and $\hat X$ are Pauli operators, $J$ is the exchange coupling and $g$ the transverse field. This model has a groundstate quantum phase transition at $g/J=1$ and a dynamical quantum phase transition when a groundstate prepared on one side of the critical point (say $g/J>1$) is evolved with a Hamiltonian on the opposite side (say $g/J<1$). Though an exact solution can be obtained using a Jordan-Wigner transformation, much remains to be understood and its dynamical and thermalisation properties are the subject of current research interest\cite{banuls2011strong,hallam2019lyapunov,azad2021phase}.
In the following, we demonstrate that the groundstate and dynamical properties of this model may be fruitfully investigated on current  quantum devices using quantum circuit MPS.

 \subsection{Groundstate Optimisation}
 \label{sec:Groundstate Optimisation}

We approach this problem using one-dimensional sequential quantum circuits\cite{banuls2008sequentially} inspired by matrix product states\cite{schollwock_review,mps_representations,orus_review,Smith2022crossing,barratt2021parallel,Lin2021real}. Fig.\ref{fig:GroundstateCircuits} illustrates the structure of these states and their representation on Rainbow. In this work we focus on bond order $D=2$\cite{schollwock_review,mps_representations,orus_review} described by a staircase of two-qubit unitaries, $U$. Higher bond orders can be achieved with an efficient shallow circuit representation\cite{Smith2022crossing,barratt2021parallel}. A crucial feature of these circuits is that although the fully translationally invariant state is infinitely deep and infinitely wide, local observables can be measured on a finite-depth, finite-width circuit with an additional unitary $V\equiv V(U)$ describing the effect of distant parts of the system on the local observables. This additional unitary is determined by a set of auxillary equations --- discussed in the Methods section and shown in Fig.\ref{fig:FixedPointCircuits} --- that we solve on chip. We optimise this fixed-point equation together with measurement of terms in the Hamiltonian in order to find the groundstate of the quantum Ising model.

Error mitigation is essential for these algorithms. We deploy a number of strategies. Foremost amongst these is choosing the best qubits. Individual qubit and gate noise can vary dramatically on current devices. In our energy optimisation experiments, we run all of the circuits shown Fig.~\ref{fig:GroundstateCircuits} in parallel on an optimally chosen set of qubits on Rainbow.
Next, we account for measurement errors and biases using a confusion matrix deduced from measurements on each of our chosen qubits. Finally, we utilise a Loschmidt echo to account for depolarisation, by implementing a circuit and its Hermitian conjugate to deduce the rescaling that depolarisation induces\cite{LoschmidtRescaling1,arute2020observation}.


%
\begin{figure*}[t]
\includegraphics[width=0.7\textwidth]{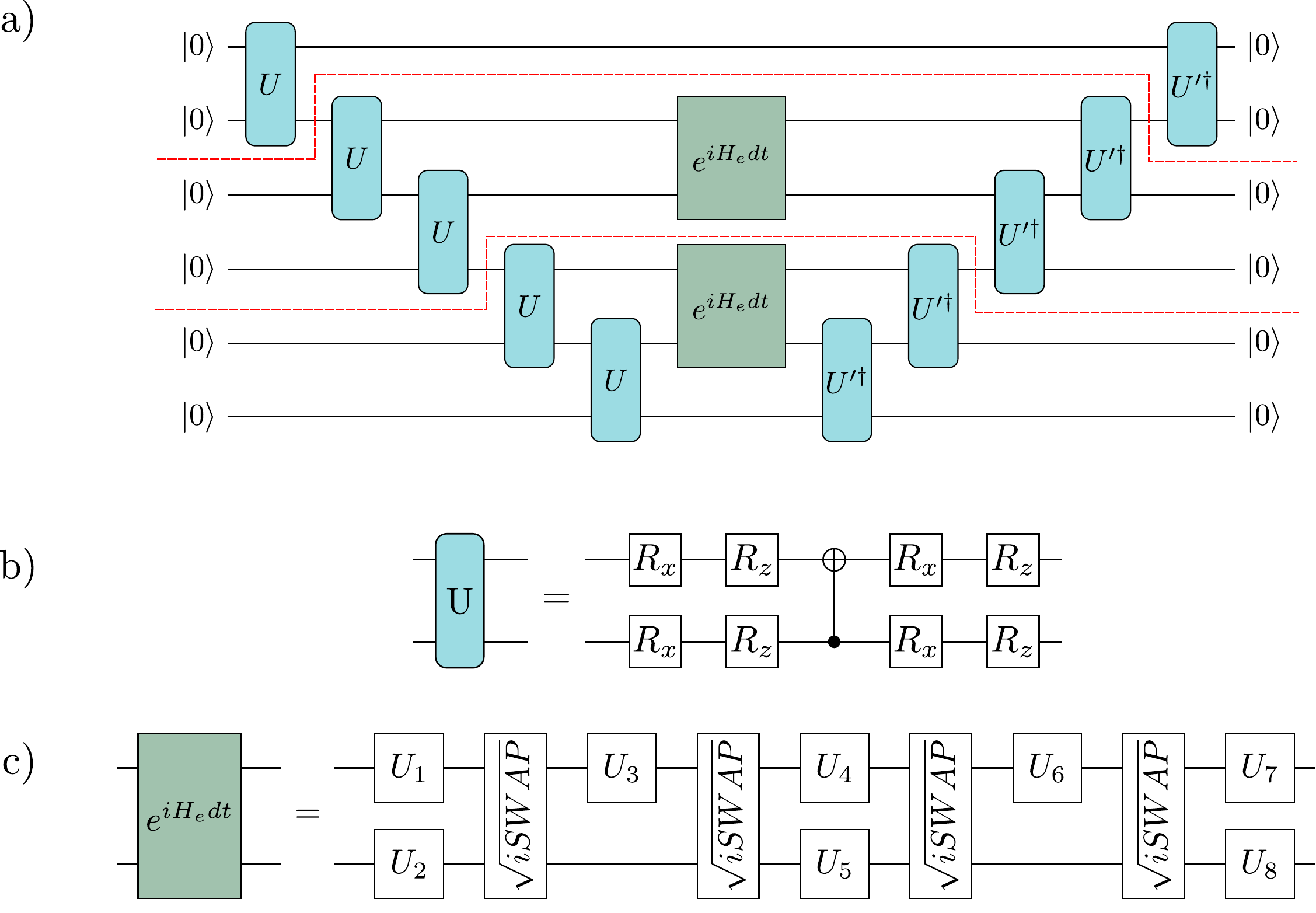}
\caption{{\bf Time-Evolution Circuits}.
a) {\it Time-evolution circuit:} The probability of measuring $|0 \rangle^{\otimes N}$ at the output, after post-selecting on the top two qubits on the right hand side, gives an approximation to $\lambda^2$ --- the square of the principal eigenvalue of the transfer matrix (indicated by the dashed red lines). This circuit provides a cost function whose optimisation over $U'$ gives the state at time $t+dt$ after starting at time $t$ with the state parametrised by $U$.
b) {\it Factorisation of the MPS and Time-evolution unitaries:}  The unitary $U$ describing the iMPS quantum state of the system is parametrised on the circuit as shown. This is reduced from the full parametrisation of a two qubit unitary in order to enable shallower circuits. There is an additional redundancy of the first $z$-rotation on the reference qubit state $|0 \rangle$ and two further angles contained in the parametrisation do not change through the dynamical quantum phase transition that we study.
c){\it Factorisation of the time-evolution unitary:} 
The two-site time-evolution unitary is factorised to the Rainbow gate set as shown. This is one of the more costly parts of the simulation in terms of circuit-depth.}
\label{fig:TimeEvolveCircuits}
\end{figure*}
\begin{figure*}[t]
\includegraphics[width=0.8\textwidth]{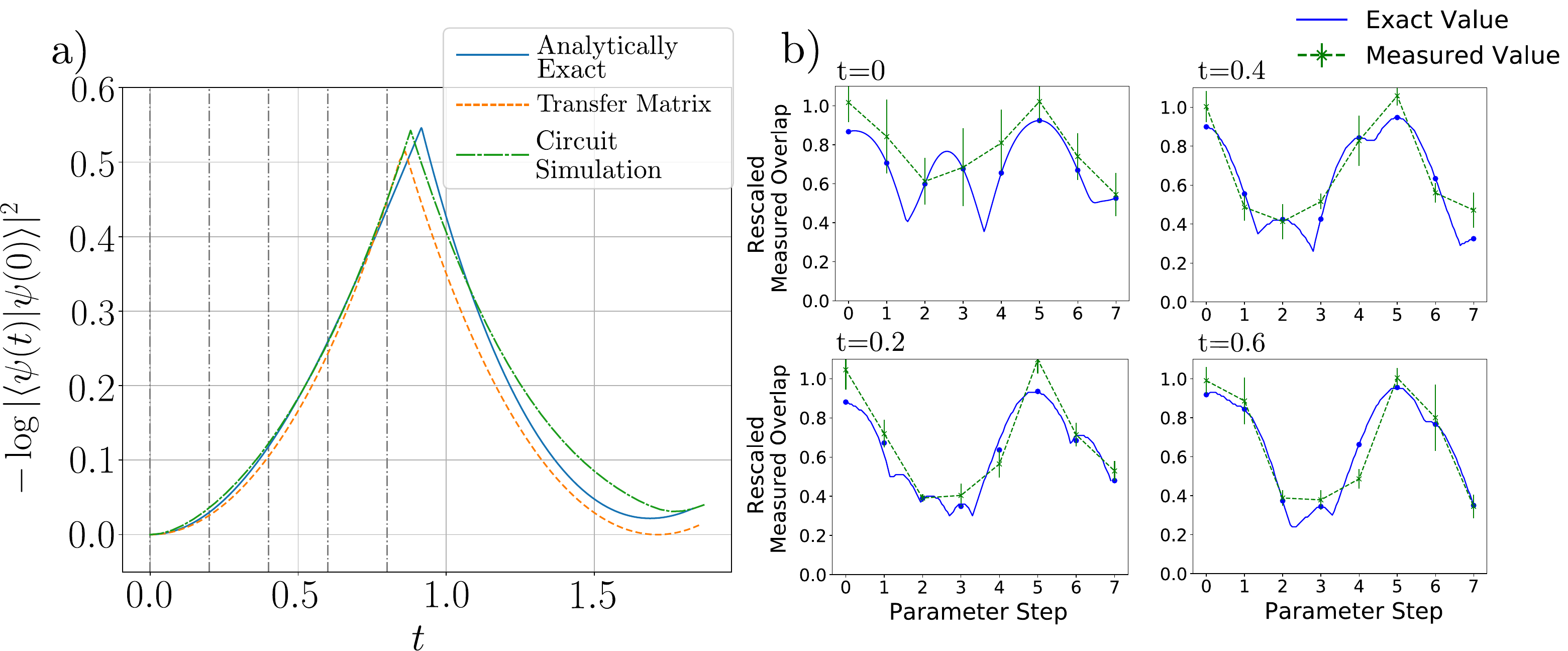}
\caption{{\bf Time-Evolution Results}.
a) {\it Dynamical Quantum Phase Transition in the Quantum Ising Model:}  The dynamics of the transverse field Ising model can be obtained analytically and form a good basis for the comparison with our quantum circuits. These are shown in blue and compared with dynamics obtained from numerically exact optimisation of the principal eigenvalue of the transfer matrix within our ansatz shown in orange, and the numerical optimisation of the circuits shown in Fig.\ref{fig:TimeEvolveCircuits} in the absence of noise shown in green. 
The results show that the circuit cost function can faithfully track the dynamics.
b){\it Cost-function evaluated on Rainbow:} The cost function evaluated along a linear interpolation in the 8 parameters from $U$ and extending through the exact update is shown for each of the indicated time-steps in a). The cost-function is rescaled by a Loschmidt echo obtained by overlapping the left hand of the circuit in a), including the time-evolution operator, with its Hermitian conjugate. The optimum value of the rescaled circuit coincides with that calculated in a classical simulation without error.}
\label{fig:TimeEvolveResults}
\end{figure*}

{\it Optimisation:} We optimise the sum of the measured quantum Ising Hamiltonian given by the circuits in Fig.\ref{fig:GroundstateCircuits}a, and the circuits  in  Fig.\ref{fig:GroundstateCircuits}b that impose consistency between the unitary $U$ defining the quantum state and the unitary $V(U)$ describing the effect of the rest of of the system on local measurements. The full cost function then is $\langle {\cal H} \rangle+Tr (\hat \rho_L-\hat \rho_R)^2$, where the second term is the trace distance between the left- and right-hand sides of the fixed-point equation first reported in Ref.\cite{barratt2021parallel} and shown in Fig.\ref{fig:FixedPointCircuits} in Methods. We carry out this optimisation using the simultaneous perturbation stochastic approximation (SPSA)\cite{spall1992multivariate}. This optimisation strategy works well deep in either phase of the quantum Ising model. In order to approach the quantum critical point, we use a quasi-adiabatic method gradually changing the Hamiltonian parameters toward those at the quantum critical point in steps, using the optimised ansatz parameters at the preceding step to start the optimisation.

{\it Results for Quantum Ising Model:}
The results of applying these methods on Rainbow are shown in Fig.\ref{fig:GroundstateResults}. These results demonstrate that, with appropriate rescaling to allow for depolarisation errors, the measured and optimised energies are close to the exact value for our ansatz even at the quantum critical point. The deviations from the analytically exact results arise primarily because we have used a reduced parameterisation of our two qubit unitaries. Fig.\ref{fig:GroundstateResults} b) shows a typical optimisation curve. A marked oscillation of unknown origin on a time-period of about half an hour was present in all of our experiments. It was particularly marked at $g=0.4$ (see Fig.\ref{fig:GroundstateResults} c) and is likely responsible for the larger error in energy at this value. 

For the quantum Ising model, $D=2$ does remarkably well and the improvement in going to $D=4$ is below the resolution of our experiments. Other models such as the antiferromagnetic Heisenberg model show a larger improvement in optimised energies by increasing bond order and are within the resolution of future experiments.

\subsection{Calculating and Optimising Overlaps}
\label{sec:Overlaps}

Before turning to quantum dynamics, we first discuss a key building block --- the overlap of translationally invariant states. For states parametrized by circuit unitaries $U_A$ and $U_B$ this overlap is given by an infinitely wide and deep version of the circuit shown in Fig.\ref{fig:StateOverlap}a. It is formally zero. However, the distance between states can be quantified by the rate at which this overlap tends to zero as the length of the system, $N$, is taken to infinity. This occurs as $\lambda^N$, where $\lambda$ is the principal eigenvalue of the transfer matrix, $E_{U_A,U_B}$, indicated by the part of the circuit contained between the dashed red lines in Fig.\ref{fig:StateOverlap}a. 

Several methods can be used to determine $\lambda$. 
Ref.\cite{barratt2021parallel} uses a variational representation of the bottom and top eigenvectors $B$ and $T$ of the transfer matrix\footnote{These correspond to the left and right eigenvectors in the usual MPS notation, were the circuit is conventionally rotated $90 \deg$ clockwise} . This involves solving fixed-point equations for $B$ and $T$ akin to those used in to calculate the expectations of the Hamiltonian in Section \ref{sec:Groundstate Optimisation}. 
Here, we use a different approach invoking the power method.  $\lambda$ is given by
\begin{equation}
\lambda 
=
 \lim_{n \rightarrow \infty} \frac{\tilde B \cdot E^n_{U_A,U_B} \cdot \tilde T}{\tilde B \cdot E^{n-1}_{U_A,U_B} \cdot \tilde T}
 =
  \lim_{n \rightarrow \infty} \frac{C_n(U_A,U_B)}{C_{n-1}(U_A,U_B)},
\label{eq:LambdaEqn}
\end{equation}
for approximations, $\tilde B$ and $\tilde T$, to the eigenvectors of the transfer matrix,
This result converges exponentially quickly with $n$, and can be particularly accurate when a good starting approximations  to $\tilde B$ and $\tilde T$ are chosen. Indeed, when either $\tilde B $ {\it or} $\tilde T$ are exact, then Eq.(\ref{eq:LambdaEqn}) converges at $n=1$. 
Fig.\ref{fig:StateOverlap}a corresponds to $n=6$ and the choices $\tilde B = {\bm I}$ and $\tilde T =| 0 \rangle  \langle 0 |$.

The results of calculating the overlaps on Rainbow in this way are shown in Fig.\ref{fig:StateOverlap}b and c. In order to correct for depolarisation errors, we divide $C_n(U_A,U_B)$ by a Loschmidt echo $C_n(U_A,U_A)$, whose value is one in the depolarisation-free case. 
Fig.\ref{fig:StateOverlap}c shows $C_n(U_A,U_B)$ and $C_n(U_A,U_A)$ calculated for different $n$. The depolarisation effects become larger as the order $n$, and consequentially depth of circuit, increases.
Fig.\ref{fig:StateOverlap}b shows how the Loschmidt-corrected results converge to $\lambda$ with $n$. By $n=4$ the estimate for $\lambda$ has converged to within error bars (aside from $n=6$ which appears to arise due to an overestimate of the Loschmidt echo. Correcting for this with an interpolated value of the Loschmidt echo brings the estimate within errors). 
The optimum balance between convergence of the power method and increasing circuit errors occurs at $n=4$ or $5$. 
Finally, Fig.\ref{fig:StateOverlap}d shows the result of optimising the overlap by varying the parameters of $U_B$ using SPSA. 

 \subsection{Quantum Dynamics}
 \label{sec:QuantumDynamics}
This method of computing overlaps can be used to time-evolve a quantum state $|\psi(U(t) \rangle$ parametrised by $U(t)$ at time $t$ to time $t+dt$ according to\cite{Lin2021real,barratt2021parallel,Barison_2021,berthusen2022quantum}
\begin{equation}
U(t+dt) = \argmax_{W} | \langle \psi(W)| e^{i {\cal H} dt } | \psi(U(t)) \rangle.
\label{eq:TimeEvolution}
\end{equation}
As in the case of direct overlaps between states, the overlap in Eq.(\ref{eq:TimeEvolution}) decays exponentially with the system size according to the principle eigenvalue of the transfer matrix. We identify circuits that approximate this principle eigenvalue as cost-functions for time-evolution. 

Just as for finding groundstates, the key to operating time evolution algorithms on NISQ devices is the management and mitigation of errors. A balance must be found between the theoretical, error-free accuracy and the effect of errors incurred on a real device.
We introduce time-evolution circuits for quantum iMPS (dramatically simplified compared to previous proposals\cite{barratt2021parallel}) that permit careful tradeoffs in implementation to enable quantum dynamics to be faithfully tracked. 

 %
The first such trade-off concerns the time-evolution operator, which must be expanded using a Trotterisation procedure\cite{trotter1959product,SUZUKI1993232,childs2019theory}. Higher order Trotterizations improve the scaling of errors with the time-step $dt$, but require deeper circuits that are more exposed to gate infidelity. Moreover, errors incurred in stochastic optimisation  may favour a larger time-step, offsetting increased resolution of the cost function against an increase in Trotter errors. 
In practice, the time-evolution operator  is the deepest part of our circuit. In order to keep this depth to a minimum we use a trick appropriate to translationally invariant states evolving with nearest-neighbour, translationally-invariant Hamiltonians. The circuit Fig.\ref{fig:TimeEvolveCircuits}a uses just the even-bonds of the Hamiltonian, but by dint of the projection back to translationally invariant states incurs errors at higher order in $dt$ than expected with a naive accounting of Trotter errors\footnote{This can be seen by noting that the time-dependent variational principle equations for evolving a translationally invariant state with just the even- or odd-bond parts of the Hamiltonian are identical to evolving using the full Hamiltonian divided by two\cite{haegemanTDVP}.}

The transfer matrix is indicated by the part of the circuit in Fig.\ref{fig:TimeEvolveCircuits}a between the red dashed lines. Time-evolution cost functions can be obtained using the power method to approximate the principal eigenvalue of this transfer matrix. The circuit show in Fig.\ref{fig:TimeEvolveCircuits}a contains two powers of the transfer matrix, together with an approximation to the bottom fixed-point of the transfer matrix equal to the identity and of the top fixed point constructed from a contraction of $U$ with ${U'}^\dagger$ with post-selection over the top two qubits. These approximations to the top and bottom fixed points are accurate to $O(dt^2)$ so that overall --- in the absence of errors --- the circuit Fig.\ref{fig:TimeEvolveCircuits}a gives an approximation to the square of the eigenvector of the transfer matrix to $O(dt^2)$. Cost functions constructed at different orders of the power method and with different approximations to the fixed points are discussed in the Supplementary Materials. 

In addition to these strategies for managing errors and making a good choice of qubits, we also require a degree of error mitigation. We average of the results obtained from four copies of the time evolution circuits: running two circuits in parallel and repeating to a total of four circuits. A Loschmidt echo is used to to mitigate depolarization errors; we deduce a rescaling by taking the circuit in Fig.\ref{fig:TimeEvolveCircuits}a setting $U'=U$ and including two copies of the time-evolution unitaries, the second with reversed time step. In this way, the circuit has similar structure to our target circuit, but has a theoretical overlap of unity. 

{\it Results:}
The results of simulation with the circuit Fig.\ref{fig:TimeEvolveCircuits}a are shown in Fig.\ref{fig:TimeEvolveResults}a and b. These results comprise two parts: a) a demonstration that circuit Fig.\ref{fig:TimeEvolveCircuits}a captures the true dynamics when run in the absence of noise; and b) measurement of the time-evolution cost-function on the Rainbow chip demonstrating. Crucially the optimum value of the measured cost function  --- measured for each time-step in the evolution along a linear interpolation from the initial parameters through the optimum --- is in the correct place. {\it In this sense, the time-evolution shown in} Fig.\ref{fig:TimeEvolveResults}a {\it is the output that would be obtained by full stochastic optimisation on the Rainbow chip}. 

The dynamical quantum phase transition studied in Fig.\ref{fig:TimeEvolveResults}a is a stringent test of our time-evolution algorithm. Starting with the parametrised groundstate at $g=1.5$ and evolving with $g=0.2$, the logarithm of the overlap of the initial state with the time evolved state, $- \log |\langle \psi(0) | \psi(t) \rangle |$, shows periodic partial revivals (corresponding to the minima in the plot in Fig.\ref{fig:TimeEvolveResults}a and dynamical quantum phase transitions (corresponding to the cusps in the plot in Fig.\ref{fig:TimeEvolveResults}a ). Observing these features requires a delicate cancellation of phase coherences in the wavefunction. 
Our results demonstrate that the Rainbow device is capable of accurately capturing the subtle features of quantum time-evolution. 

%% file: Discussion.tex
We have demonstrated that tensor network methods endow NISQ devices with the power to simulate the groundstate and dynamics of quantum critical systems. Such systems pose amongst the greatest challenge for classical simulation techniques and provide a forum with genuine potential for quantum advantage. 
Our algorithms operate directly in the thermodynamic limit - avoiding difficulties of finite-size scaling near criticality.
We demonstrate that translationally-invariant variational approximations to the groundstate can be optimised on current devices to good accuracy even at the quantum critical point, and present a cost function for time-evolution--- a significant simplification over previous proposals --- that can faithfully track dynamics when implemented on the Rainbow device. 

An explicit, stochastic optimisation of this cost function remains a subject for future study. There are many different schemes available for such optimisations and choosing an appropriate one  is an important task. Moreover, sampling costs of the combined measurement of overlaps and stochastic optimisation of the updated state ansatz must be kept to realistic levels. Offsetting the accuracy with which the overlap of any particular trial update is determined against the number of potential updates that are tested provides a further space for optimisation. The quantum circuit effectively provides stochastic corrections to a classical model guiding the choice of test updates and poses an interesting problem in quantum control\cite{vandersypen2005nmr,dong2010quantum,bukov2018reinforcement}. Ultimately, the task is one of performing tomography of the updated state to an MPS approximation\cite{Lanyon_2017}.

Effective error mitigation is crucial in using NISQ devices for quantum many-body simulation, and form a central part of our discussion. Tensor networks permit a systematic tradeoff between the error-free accuracy of circuits, and the errors incurred due to circuit infidelity. 
For example, the time-evolution circuits present a suite of refinement parameters --- such as bond-order, Trotterisation order and timestep, order of power method --- that can be chosen to optimise performance on a given device together with a confusion matrix to correct for measurement errors and a Loschmidt echos to correct for depolarisation. 
Other strategies can be deployed as the accuracy of simulations improves. Perhaps the most exciting follow the structure of the tensor networks themselves; averaging over the gauge freedom intrinsic to the auxilliary space of the tensor network states is natural first step. Combining with the matrix product operator encoding of errors discussed in Ref.\cite{guo2022quantum} has potentially only constant overhead in circuit depth if the staircase of state and MPO unitaries are aligned (in a similar manner to the interferometric measurement of topological string order parameters in Ref.\cite{Smith2022crossing}).

Extensions include using higher bond order to refine simulation accuracy. In the present case, improvements expected from increasing the bond order fall below the resolution of our experiments. Other problems (such as the antiferromagnetic Heisenberg model) show a greater improvement with bond order and may be feasible. Though rastering one-dimensional MPS algorithms over a higher-dimensional system is competitive in classical applications, in the quantum setting, it may be favourable to take more direct advantage of the connectivity of the quantum device. Our code can be modified to accommodate the two-dimensional sequential circuits proposed in Ref.\cite{banuls2008sequentially}, and the isometric two-dimensional algorithms of Ref.\cite{zaletel2020isometric} are another promising avenue. 

The use of machine learning in many-body quantum physics is intriguing recent development\cite{carrasquilla2017machine,rodriguez2019identifying,ch2017machine,broecker2017machine}. 
One use has been identifying phases in classical simulation ({\it e.g.} monte carlo simulations) of many-body quantum systems, thereby potentially enhancing the ability of these established approaches to reveal new physics. In its quantum application, such an approach might simultaneously mitigate intrinsic errors in the simulation and errors due to the infidelity of the device\cite{herrmann2021realizing}. Combining with tensor network realisation of the machine learning components\cite{stoudenmire2016supervised,huggins2019towards} may enable a compact realisation of these combined aims. Indeed, combining the methods of Ref.\cite{guo2022quantum} with our scheme would constitute precisely this.

Tensor networks provide a systematic way to structure quantum simulations on NISQ devices. Our results show how they can be used to analyse quantum critical systems in the thermodynamic limit. There are many promising ways in which their application can be extended.

%% file: MethodsA.tex
\subsection{Fixed-point equations}
\begin{figure}[ht]
\includegraphics[width=0.45\textwidth]{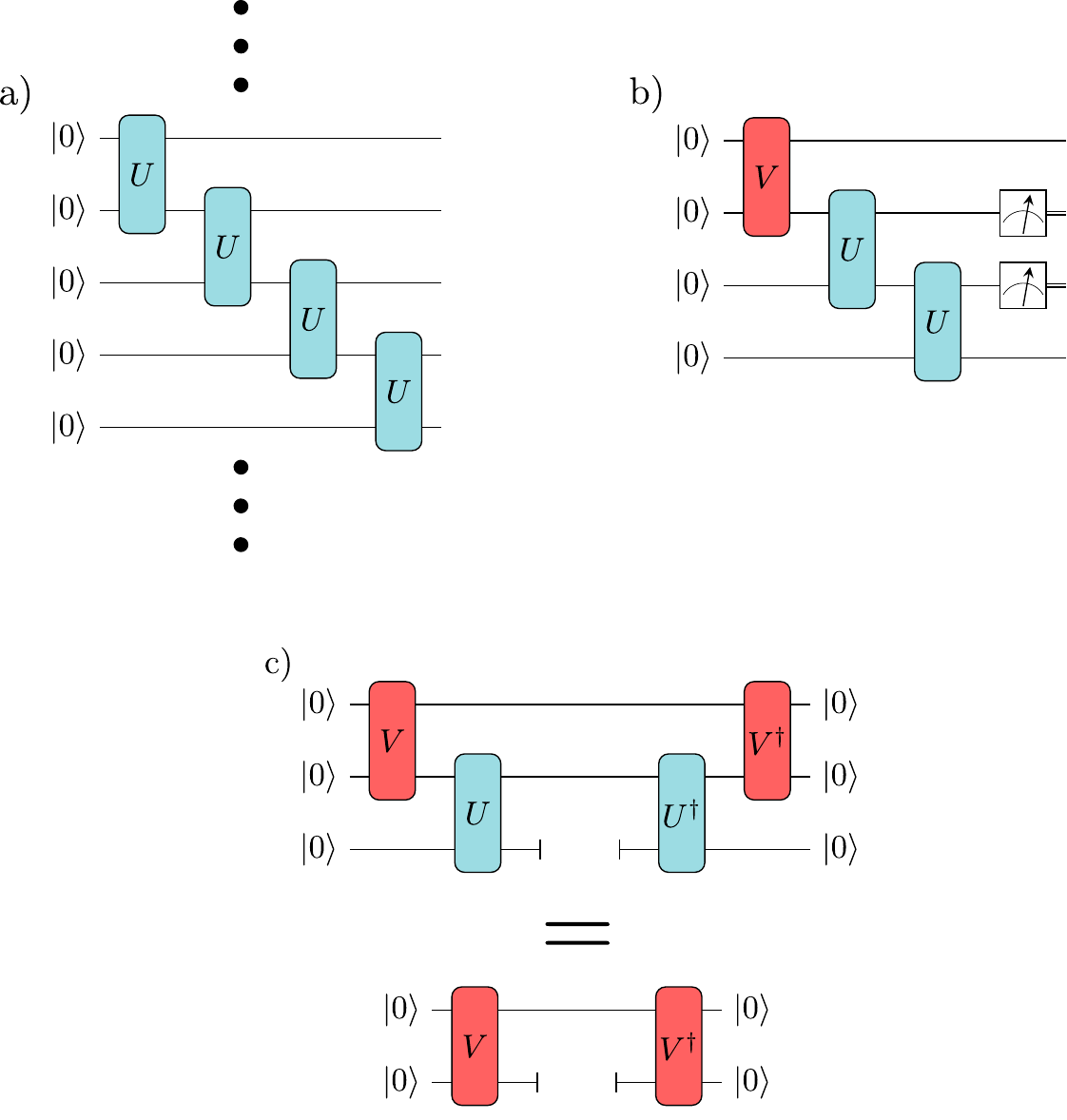}
\caption{{\bf Finite Circuit Representation of iMPS states} a) The iMPS state is formally represented by an infinitely wide and deep quantum circuit. b) Local observables can be represented by a finite circuit by introduicing the environment tensor $V \equiv V(U)$. c) $V$ is determined by the fixed-point equations shown, which may be evaluated on circuit.}
\label{fig:FixedPointCircuits}
\end{figure}

A translationally-invariant MPS formally requires an infinitely wide, infinitely deep quantum circuit to represent it (see Fig.\ref{fig:FixedPointCircuits}a). As demonstrated in Ref.\cite{barratt2021parallel}, local observables of this translationally invariant state can be measured on a finite circuit by introducing the environment tensor $V$ which summarises the effects of distant parts of the quantum state on the local measurement (see Fig.\ref{fig:FixedPointCircuits}b ). This tensor is determined as a function of the state unitary, $U$, by solving the fixed point equations shown in Fig.\ref{fig:FixedPointCircuits}c). 
The circuits implemented in Fig.\ref{fig:GroundstateCircuits} correspond to the various terms involved in the trace distance between circuits on the left- and right-hand side of the fixed point equation, Fig.\ref{fig:FixedPointCircuits}c: $Tr(\hat \rho_L-\hat\rho_R)^2$. There are three terms corresponding to the trace norm of each side of the equation and the cross terms between them. 
In practice, when optimising the energy, the cost-function for the fixed-point equations and the various contributions to the energy are optimised simultaneously using SPSA.

\label{sec:FixedPoint}

%% file: MethodsB.tex
\subsection{Error Mitigation Strategies}
\label{sec:ErrorMitigation}

We deploy a number of error mitigation strategy in order to obtain our results, summarised as follows:

{\it Qubit Choice:} The measured single and two qubit gate fidelity vary across the Rainbow device, and vary over time on each qubit. We can reduce the error in circuit executions by picking connected sets of qubits that have the lowest error rates. This is best done empirically by running a reference circuit across the target qubits and choosing the qubits that have the highest fidelity with the known target state. To best capture the effects of noise on the circuits that we are using, we apply an MPS circuit and then the Hermitian conjugate of the same MPS circuit. We measure the probability of returning to the initial all-zeros state after $N$ applications of this Loschmidt echo circuit and use this to determine the best qubit sequences to use. We also employed qubit averaging where the same circuit was run on multiple non-intersecting sets of qubits in parallel and the results averaged. 

{\it Measurement Errors:} The readout of quantum states on the Rainbow device,  can be  biased towards certain bit strings. To correct for this we learn how the readout is biased by measuring states that produce fixed bit-string outputs. We encode any deviations in the known bit strings by a \emph{confusion matrix}. 
This can be inverted and used to correct the 
measurement bias. The size of the confusion matrix is $2^{M}$ where $M$ is the number of measured qubits. 
For all circuits in this work where this error mitigation strategy is applied the number of measured qubits is only 2. 
In principle, this method may not be scalable for quantum MPS methods. 
The size of the confusion matrix needed to correct the fixed point calculation circuits grows polynomially with the bond dimension of the MPS, and can potentially erode any performance benefits from running MPS simulations on a quantum device.
However, where we are interested in the relative rather than absolute values of measured circuits --- such as in our time-evolution and fixed-point circuits --- this bias can be neglected without adversely affecting results. 

{\it Depolarisation} is a major source of error in the measured outputs of our quantum circuits. We correct for this using a Loschmidt echo in two different forms for our energy optimisation and time-evolution circuits\cite{LoschmidtRescaling1,arute2020observation}. In the case of optimisation of the energy of the Ising model, 
we know the energy of the ground state when the coefficient of the interacting term is set to 1 and the single site terms are set to 0. We also know the parameters in our ansatz corresponding to the ground state with these parameters. We measure the energy of this state on the device to get a rescaling and correct all future interacting terms measured on the device with this value. For time-evolution circuits,
we use the fact that the overlap of a circuit with its Hermitian conjugate should always be 1. We choose a test circuit that is representative of the complexity of our target circuits and measure its overlap with its Hermitian conjugate. Target circuits are divided by this value to allow for depolarisation. In order to reflect the complexity of the time-evolution circuits in Fig.\ref{fig:TimeEvolveResults} we use a circuit MPS state and apply the time evolution unitary followed by its inverse, before finally overlapping with the Hermitian conjugate of the circuit MPS state. 

{\it Floquet calibration}\cite{Floquet1,arute2020observation} has been developed to correct the angles on the native Rainbow two qubit gate so that desired two qubit unitaries can be applied more accurately. Floquet calibration proceeds by repeating a two qubit gate multiple times on the chip to amplify small discrepancies in the angles that are being applied. The measured deviations can then be compensated to increase the fidelity of the implemented circuit with the intended circuit. This method was found to have only a small impact on the energy measurements, and was not as effective as the confusion matrix and Loschmidt rescaling techniques. For this reason, we did not apply Floquet calibration to our final calculations of energy or time evolution circuits. 

%